# Best practices for the manual curation of Intrinsically Disordered Proteins in DisProt

Federica Quaglia [1,2], Anastasia Chasapi [3], Maria Victoria Nugnes [2], Maria Cristina Aspromonte [2], Emanuela Leonardi [2], Damiano Piovesan [2], Silvio C.E. Tosatto [2,*]

[1] Institute of Biomembranes, Bioenergetics and Molecular Biotechnologies, National Research Council (CNR-IBIOM), Bari, Italy.
[2] Department of Biomedical Sciences, University of Padova, Padova, Italy.
[3] Biological Computation & Process Laboratory, Chemical Process & Energy Resources Institute, Centre for Research & Technology Hellas, Thermi, Thessalonica 57001, Greece.

* Corresponding author: Tel: +39-049-827 6269; email: silvio.tosatto@unipd.it

Keywords: Guidelines, biocuration, intrinsic disorder, DisProt

## Abstract

The DisProt database is a significant resource containing manually curated data on experimentally validated intrinsically disordered proteins (IDPs) and regions (IDRs) from the literature. Developed in 2005, its primary goal was to collect structural and functional information into proteins that lack a fixed three-dimensional (3D) structure. Today, DisProt has evolved into a major repository that not only collects experimental data but also contributes significantly to our understanding of the IDPs/IDRs roles in various biological processes, such as autophagy or the life cycle mechanisms in viruses, or their involvement in diseases (such as cancer and neurodevelopmental disorders). DisProt offers detailed information on the structural states of IDPs/IDRs, including state transitions, interactions, and their functions, all provided as curated annotations.

One of the central activities of DisProt is the meticulous curation of experimental data from the literature. For this reason, to ensure that every expert and volunteer curator possesses the requisite knowledge for data evaluation, collection, and integration, training courses and curation materials are available. However, biocuration guidelines concur on the importance of developing robust guidelines that not only provide critical information about data consistency but also ensure data acquisition.This guideline aims to provide both biocurators and external users with best practices for manually curating IDPs and IDRs in DisProt. It describes every step of the literature curation process and provides use cases of IDP curation within DisProt.

Database URL: https://disprot.org/

## Introduction

Intrinsically disordered proteins (IDPs) and intrinsically disordered regions (IDRs) are key players in a plethora of biological processes, wielding their unique structural flexibility to participate in vital cellular functions [1]. Their lack of a stable three-dimensional structure under physiological conditions, challenging traditional structural paradigms, prompts the necessity for specialised biocuration efforts. The DisProt database stands as a gold standard resource in this endeavour, collecting manually curated experimental data that describe the multifaceted roles of IDPs and IDRs [2–5]. The landscape of intrinsically disordered proteins spans critical domains such as viral processes, autophagy, and disease pathways, including cancer. Their ability to rapidly adapt their conformation allows them to engage in multiple interactions, modulating cellular responses [6,7]. As their relevance becomes more evident, the need for precise, comprehensive, and reliable biocuration gains increasing importance.

Biocuration, the curation of biological information, plays an instrumental role in capturing the several aspects that characterise IDPs and IDRs. The DisProt database has etched itself as a pivotal resource, gathering an expansive collection of over 2400 entries across diverse species and biological kingdoms. Expert biocurators sift through the scientific literature to gain insight into structural states, transitions, interactions, and functions associated with IDPs and IDRs [3].
In this context, this guideline aims to illuminate the best practices for biocurators but also extends its reach to external users that seek comprehensive insights into the world of IDP and IDR curation. By dissecting the intricate steps of literature curation and offering real-world use cases, this guideline fosters a comprehensive understanding of the curation process and underscores the importance of DisProt in advancing the knowledge of IDP biology. The guideline spans the landscape of IDP and IDR curation, encompassing all aspects of curation processes, i.e. data prerequisites, structural ontologies, literature retrieval strategies, functional annotations and submission procedures. Moreover, by exemplifying the curation of specific well-known IDPs, such as ATG8-interacting protein 2 and RAF proto-oncogene serine/threonine-protein kinase, this guideline showcases the various aspects of IDP and IDR curation within the DisProt framework.

## Overview of the IDP/IDR manual curation process

The manual curation process in DisProt, summarised in Figure 1, can start from one of the following two key methods: a candidate publication or a candidate protein.
If curation begins with a candidate protein, a publication that characterises the protein's disordered nature must be identified using PubMed [8] or EuropePMC [9].
If curation begins with a candidate publication, the proteins described in the publication must be identified. Regardless of the starting point, once a protein and a relevant publication have been chosen, the actual curation process is the same (as outlined in Figure 1), which involves extracting the intrinsic disorder-related information from the chosen publication.

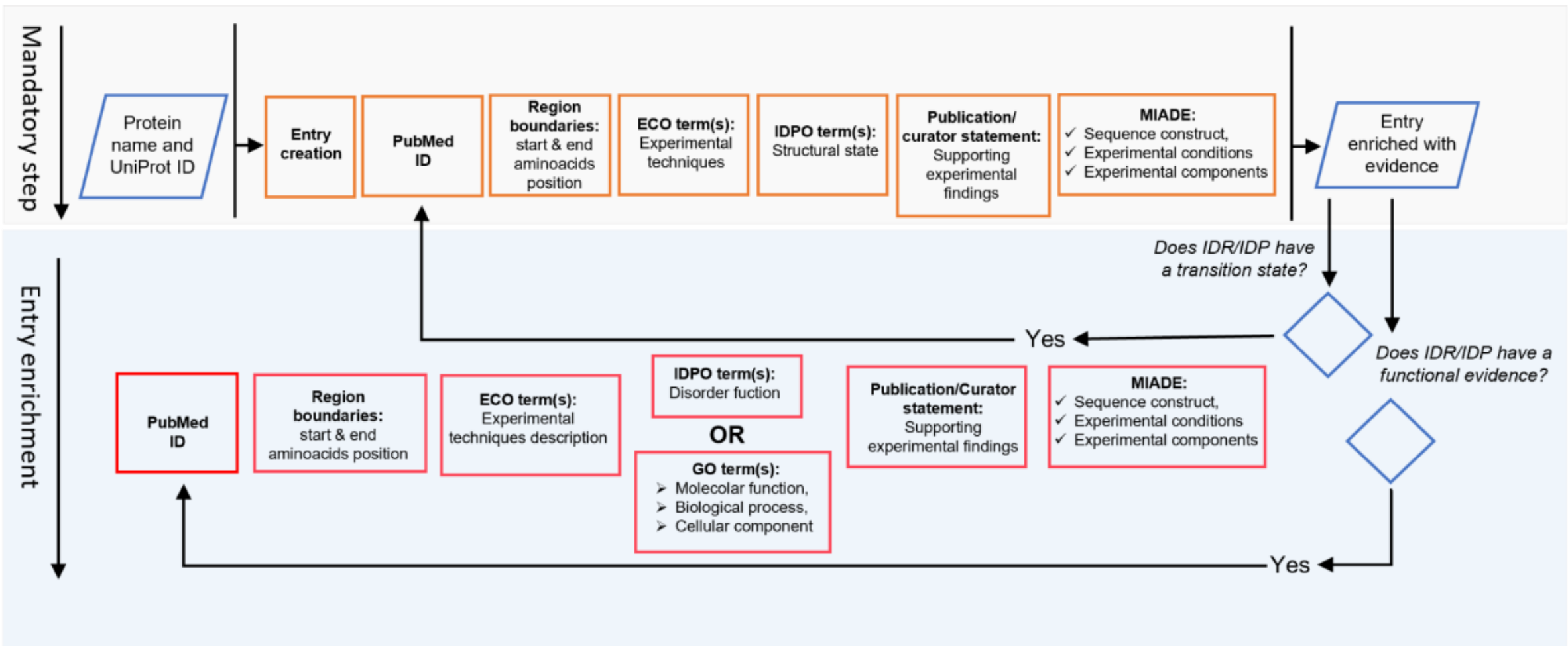

**Figure 1.** Workflow describing the important steps in the DisProt curation process. Abbreviations, ECO: Evidence and Conclusion Ontology; IDPO: Intrinsically Disordered Proteins Ontology; MIADE: Minimum Information About Disorder Experiments; GO: Gene Ontology

As outlined in the workflow, it is crucial to emphasise that manual curation of a novel IDPs or IDRs (with at least 10 disordered residues) requires the following essential information:

- A **peer-reviewed article** detailing the IDP/IDR, supported by reported scientific experiments.
- The identification of the protein's **UniProt ID** mentioned in the publication paying attention to the *Organism*.
- At least one **experimental method** defining the experimental setup used to generate the annotated information.
- The definition of the protein **region boundaries** expressed as positions (start-end) in the corresponding UniProt entry.
- The annotation of the **intrinsic disorder structural state** using IDPontology.

- A **publication statement** consisting of sentences extracted from the publication, which provides information supporting the experimental findings and properly attributes the source of the presented information. Alternatively, a **curator statement** contains sentences provided by the curator (often experts in a particular field/method), significantly describing the information extracted from the publication and offering insight into what viewers are observing.

It is also important to specify that if an entry associated with a specific UniProt ID is already present in DisProt, a curator can add new evidence regarding its disordered state or evidence related to its functions. In this case, the curator who did not create the entry must request permission to edit it, as they do not “own” it. After defining the disorder state of a protein or region, the curator can add further evidence describing structural transitions or disorder-specific functions (Figure 1).

# Structuring data with the use of ontologies

In recent years, DisProt has been improved to facilitate structured curation with a controlled vocabulary using three ontologies:

- Intrinsically Disordered Proteins Ontology (IDPO), accessible at https://disprot.org/ontology [3], describes structural aspects, states, transitions of an IDP/IDR, as well as self-functions, and functions directly associated with their disordered state.
- Gene Ontology **(**GO) [10] (URL: http://geneontology.org/) is integrated to describe three key aspects within the biological domain related to IDP/IDR: molecular functions (MF), biological processes (BP), and cellular components (CC).
- Evidence and Conclusion Ontology (ECO) (URL: https://www.evidenceontology.org/) represents the experimental techniques used to assess the disordered structural state in a protein or related aspects.

Curators have the option to select ontological terms that best describe the information reported in the publication. For each piece of evidence that indicates the structural state, the disorder function, and the applied method, curators should utilise the provided ontologies.

# Methods for retrieving IDP-related evidence

The extraction of disorder-related information can be one of the most significant challenges. Various strategies can be employed to retrieve information regarding the disorder status and functions of these proteins:

1. Identify suitable publications that report experimental evidence of disorder state from **PubMed** and **Europe PMC**. Curating IDPs/IDRs can be accomplished by constructing a “query” using a combination of protein/gene names (or synonyms) and disorder-related keywords. Recommended keywords or effective detection of intrinsic disorder mentions in a publication include terms such as: *“(intrinsic) disorder”, “unstructured”, “unfolded”, “flexible/flexibility”, “(high) mobility”, “missing residues”, “electron density”.* Authors sometimes list all the detected or missing electron density regions in their crystal structure, but often without explicitly using the term ‘disorder’. It can also be helpful to search for terms like: "*visible*" or "*missing*," which may indicate regions with the presence or absence of structure. If the terms mentioned above are not present in the publication, it may also be useful to search for terms that refer to experimental techniques frequently used to assess intrinsic disorder, such as *“NMR”, “circular dichroism”*, *“SAXS”*, etc.

2. The second strategy involves consulting databases based on experimental methods or biological mechanisms closely associated with the disorder state. The databases mentioned in Table 1 are also cross referenced in DisProt.

| Resource | Description | URL |
|---|---|---|
| PDBe [11] | X-RAY crystallography, NMR-related methods, electron microscopy protein structures | https://www.ebi.ac.uk/pdbe |
| BMRB [12] | Spectral and quantitative data derived from Nuclear Magnetic Resonance (NMR) spectroscopic | http://www.bmrb.wisc.edu |
| SASBDB [13] | Small Angle X-ray Scattering (SAXS) and Small-Angle Neutron Scattering (SANS) experiments | https://www.sasbdb.org/ |
| EMDB [14] | Electron Microscopy Data Bank for electron cryo-microscopy, single-particle analysis, electron tomography, and electron crystallography | https://www.ebi.ac.uk/pdbe/emdb/ |
| PCDDB [15] | Circular dichroism (CD) and synchrotron radiation CD (SRCD) spectral data and their associated experimental metadata. | http://pcddb.cryst.bbk.ac.uk/home.php |
| PhaSePro [16] | Manually curated resource of proteins driving liquid-liquid phase separation (LLPS) | https://phasepro.elte.hu/ |
| AmyPro [17] | Validated amyloid precursor proteins and their amyloidogenic sequence regions | https://amypro.net/#/ |
| ELM [18] | Curated database of short linear motifs (SLiM) in Eukaryotes | http://elm.eu.org/searchdb.html |
| UniProt [19] | Resource of protein sequence and functional information | https://www.uniprot.org/ |
| MobiDB [20] | Database of protein disorder and mobility annotations | https://mobidb.bio.unipd.it/ |

**Table 1.** The table shows a list of databases that can be consulted to retrieve information and references on experimental studies in proteins. Some of these resources pertain to techniques used in structural biology and are linked to scientific articles, while others, like MobiDB, serve as sources to extract information about the possible disorder state of a protein, which can be either predicted or curated. Indeed, in the last version of DisProt, another track specifically highlights the disordered regions derived from the missing residues of the PDB, as calculated by MobiDB (consensus trace).

## Finding a UniProt identifier for a protein

The peer-reviewed article used as the basis of the curation should ideally report the UniProt accession number of the protein described. If this information is available, the curator should use the specified UniProt accession.

If this information is not available, the curator may encounter many exceptions:

- The authors use a different protein identifier (e.g. Ensembl ID, HGNC, etc.). In this case the ID mapping service of UniProt (URL: https://www.uniprot.org/uploadlists/) to retrieve the corresponding UniProt accession, is encouraged.
- The authors reference only the protein sequences. The curator should use that sequence as a query to search for a UniProt ID using the built-in BLAST function. It is advisable to restrict the search criteria as much as possible (e.g., limiting to human proteins, vertebrates, bacteria, etc.).

The results of UniProt searches should be manually assessed, and the curator should use their judgement to select the most fitting protein. In the selection of the best match, the criteria described earlier in cross-checking (protein length, source organism, region boundaries, etc) can be used. If the same sequence is present in UniProt under various accession numbers, priority should be given to SwissProt entries, if available. If the curator cannot find an available SwissProt entry, then the longest TrEMBL entry should be considered.
Discrepancies between UniProt and publication sequences can arise because of errors in the publication or modifications to the UniProt sequence after the publication date. In such cases, curators should compare the protein sequences, taking into account the provided sequence (if available from the authors), sequence length, and region boundaries.

## Identifying and annotating the correct boundaries of a protein

Discrepancies may also occur for the sequence reported in the publication compared to the official UniProt one. This following curation example (DisProt entry DP02957, URL: https://disprot.org/DP02957) shows how curators should always check the amino acid boundaries of the IDR described in the publication before annotating in DisProt.
The authors of the paper *“Monomeric solution structure of the prototypical 'C' chemokine lymphotactin.”* published in Biochemistry [21], used nuclear magnetic resonance to analyse the IDRs of the human protein Lymphotactin. The UniProt accession is P47992 (URL: https://www.uniprot.org/uniprotkb/P47992/entry). While it is not explicitly stated in the paper, the authors provide a GenBank accession (U23772) that can be mapped with the UniProt ID mapping service (URL: https://www.uniprot.org/id-mapping/) to retrieve the corresponding UniProt accession.

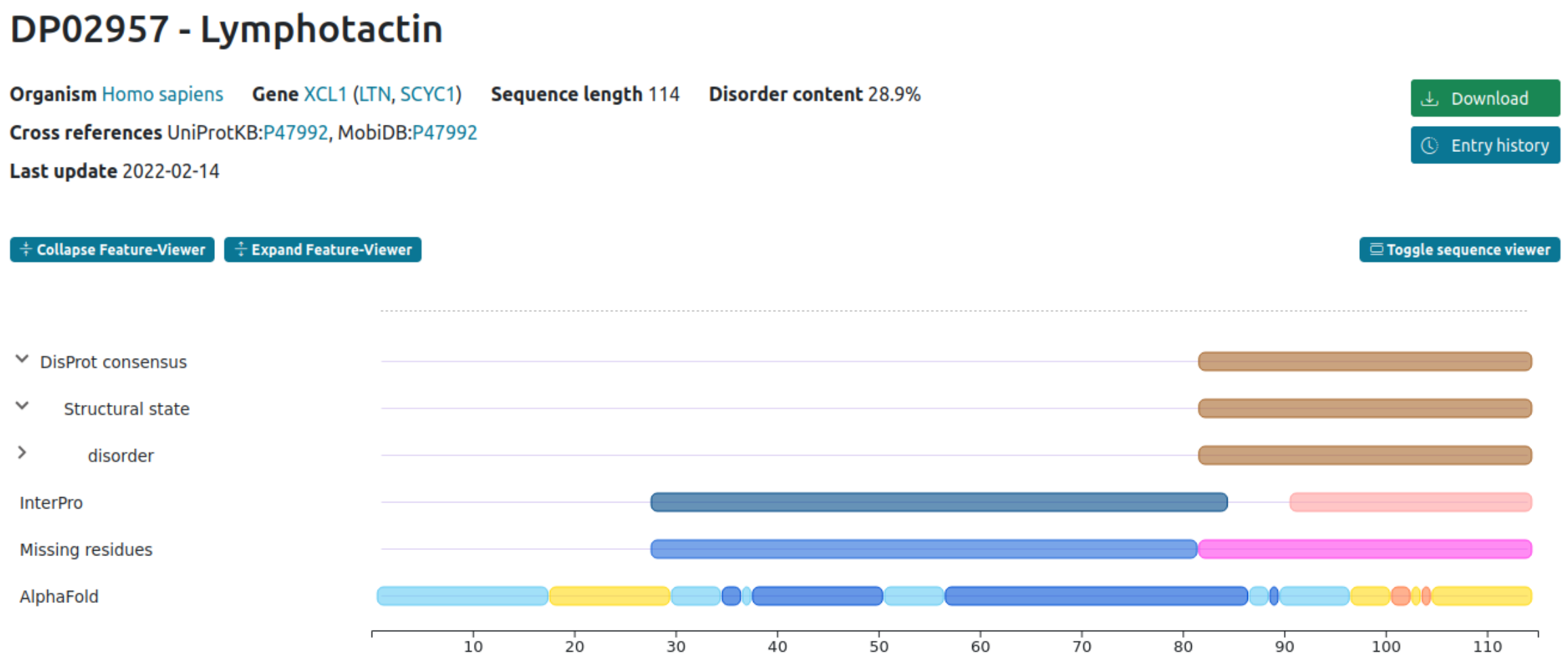

**Figure 2.** Example of the human lymphotactin protein available in DisProt.

The authors of the publication provide several statements, extracted from different article sections, that should be used when annotating the IDRs of human lymphotactin in DisProt:

- ➔ **Abstract:** “*Two regions are dynamically disordered as evidenced by 1H and 13C chemical shifts and {15N}-1H NOEs: residues 1–9 of the amino terminus and residues 69-93 of the C-terminal extension.*”
- ➔ **Results:** “*A steady decline is seen in the heteronuclear NOE values for the unstructured*

*residues approaching the ends of the N- and C-termini with negative NOE values observed for residues near each terminus; these values are consistent with large-amplitude motions on the picosecond to nanosecond time scale and a complete lack of stable secondary or tertiary structure.*"

- ➔ **Results:** "*No long-range NOEs were observed for the 9 N-terminal and 26 C-terminal residues.*"
- ➔ **Results:** "*Residues 1–8 and 69–93 are highly disordered.*"
- ➔ **Discussion:** "*Residues comprising the unique C-terminal sequence of hLtn are entirely disordered in solution (Figures 4 and 5), but residues 9–68 adopt the conserved fold observed for all other chemokines (Figure 6).*"

In this example, two crucial factors need consideration when curating the disordered regions identified by the authors. Firstly, DisProt requires a minimum of 10 residues to annotate the structural state, rendering the initial region (residues 1-8) too short for DisProt annotation. Secondly, the sequence used in the publication for defining IDRs does not align with the UniProt canonical sequence (P47992) of the protein under study. The experimental sequence in the publication corresponds to the mature secreted form of the protein, excluding the N-terminal signal sequence spanning the first 21 amino acids. To verify this, curators should refer to the "*PTM / Processing*" section of the UniProt entry (P47992) to find evidence of the signal peptide (region 1 - 21) in human lymphotactin. This confirms that the studied sequence ranges from residue 22 to 114, necessitating the annotation of boundaries in DisProt as 91 - 114, not 69 - 93.

## Defining experiments and cross-references for an IDP/IDR

The experimental technique used to analyse an IDP/IDR and its related disorder aspects can be annotated using ECO terms. When available the most specific technique must be selected using the ECO term or the technique's name (Figure 3).

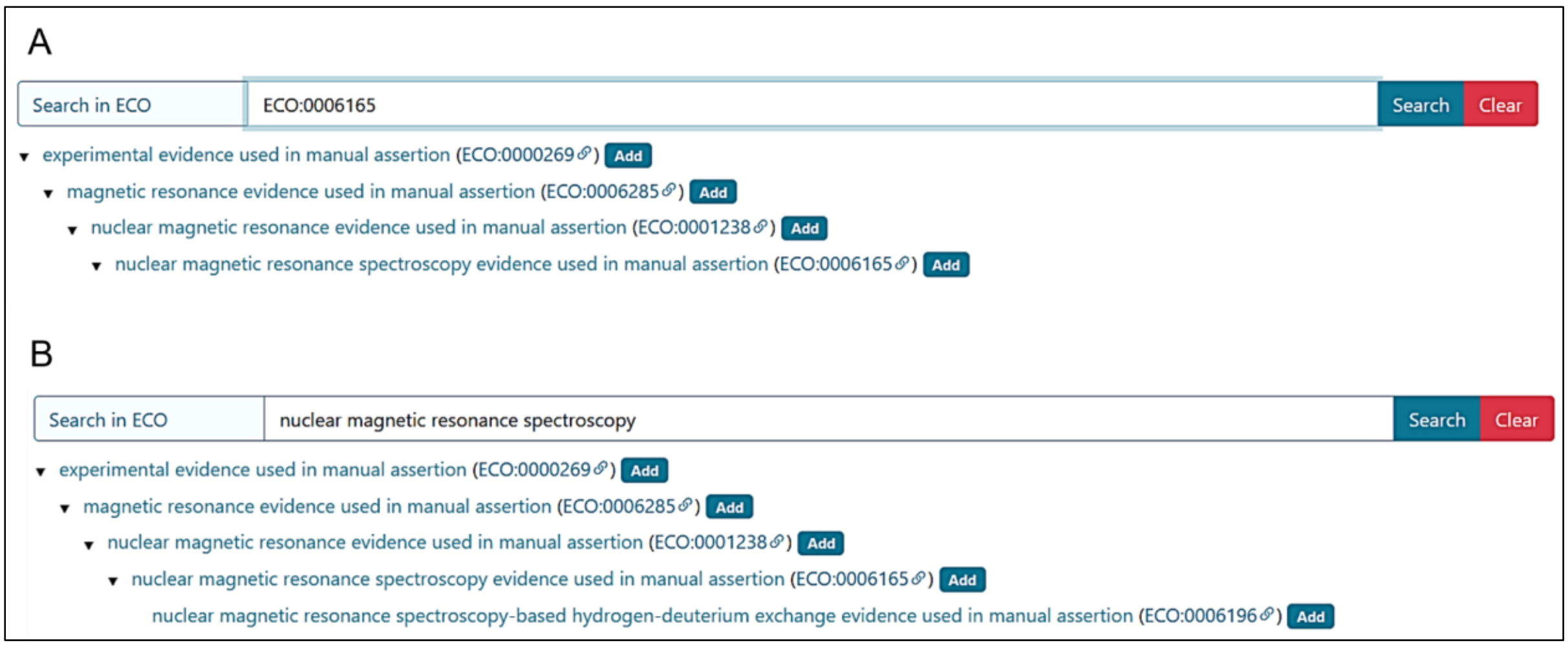

**Figure 3. Entry curation page.** Curator can retrieve the ECO technique using the ECO term (A) or the technique description (B) in the "Search in ECO" bar, press the Search button and add the appropriate method.

The curator is also encouraged to add cross-references to point to other databases, which can provide additional evidence regarding the disorder state or function. DisProt is currently cross-referenced to PDB, BMRB, PCDDB, SASBDB, EMDB, PhasePro, AmyPro, ELM. A list of source databases that can be cross-referenced in disorder-related publications, is provided in Table 2.

The system automatically suggests the cross-reference after the reference identifier is added as evidence (Figure 4). For example if the authors used X-Ray for the structure detection and deposited the structure in PDB the PDB code should be added. However, the curator should verify the information before accepting the suggestion.

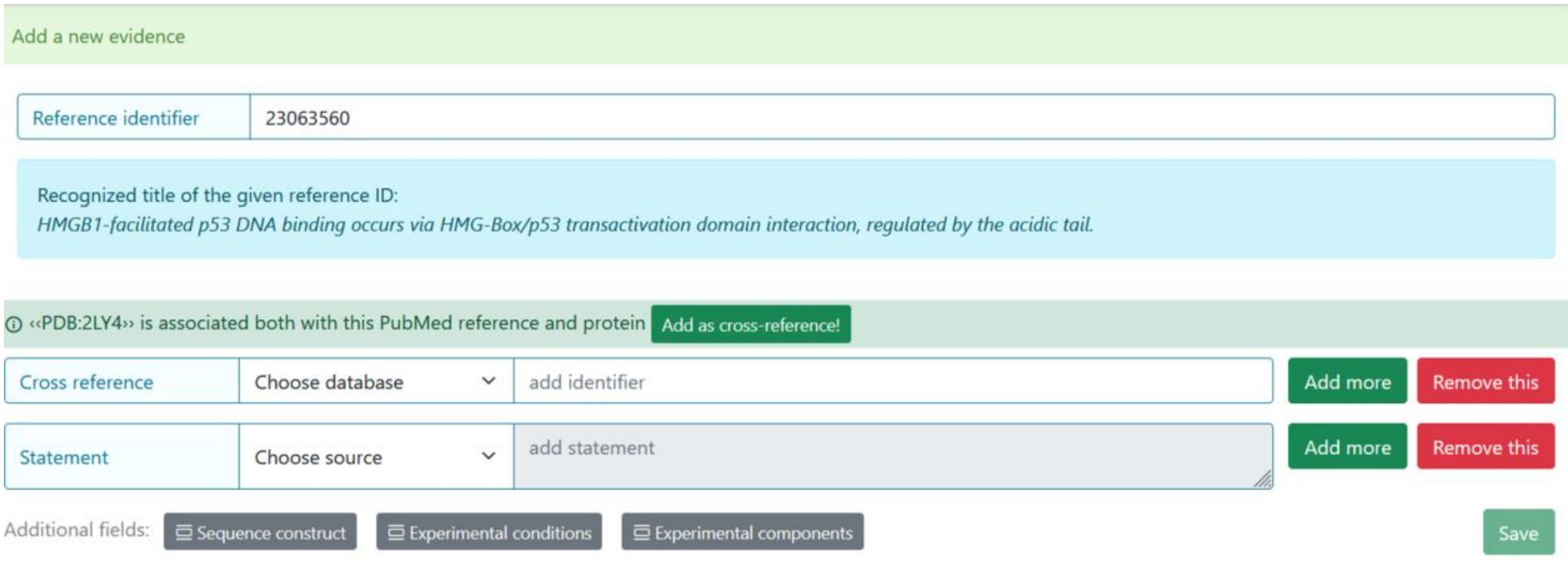

**Figure 4. Entry curation page.** Following the addition of the Reference identifier, the system automatically retrieves the PDB code related to the specified publication. Before adding it, the curator must verify that the information is correct.

# Identifying intrinsic disorder structural states and transitions

The following sections describe how to look for structural states and structural transitions pertaining to IDPs/IDRs in a publication.

## Defining types of structural states

Four structural state terms are available in the IDP ontology. There are two high level terms to define the presence or the lack of structure (**A** and **B**) and two subtypes of disorder to define a more specific structural state (**a** and **b**):

A. ***Disorder***: a non-compact state in which the protein lacks a stable three-dimensional structure in isolation, covering both secondary structural elements and tertiary structure
    a. ***Pre-molten globule*****:** a condensed but not compact state, with residual secondary structure, describing many native and non-native conformations in rapid equilibrium
    b. ***Molten globule***: a compact state, with native secondary structure but lacking specific native tertiary structure
B. ***Order***: a compact state with a stable three-dimensional structure, in which most atoms have a fixed and stable position in relation to each other

## Defining types of structural transitions

The IDP ontology includes both transitions of IDPs and IDRs into a more ordered state or to a more disordered state, as follows:

I. **transitions to a more ordered state**: *disorder to pre-molten globule, disorder to molten globule, disorder to order, pre-molten globule to molten globule, pre-molten globule to order, molten globule to order.*
II. **transitions to a more disordered state**: *order to molten globule, order to pre-molten globule, order to disorder, molten globule to pre-molten globule, molten globule to disorder, pre-molten globule to disorder*.

# Identifying functions associated with IDPs and IDRs

The following sections describe how to look for functions pertaining to IDPs/IDRs in a publication.

## Defining functions through Gene Ontology terms

Gene Ontology - the largest knowledgebase providing information on the functions of genes - can be used to describe functional aspects of an IDP or IDR. GO describes our knowledge of the biological domain with respect to three aspects:

- Molecular Function (MF), molecular-level activities performed by gene products.
- Biological Process (BP), larger processes, or 'biological programs' accomplished by multiple molecular activities.
- Cellular Component (CC), cellular location where a gene product performs a function.

More information is available from the dedicated Gene Ontology documentation page (URL: http://geneontology.org/docs/ontology-documentation/).
When annotating with Gene Ontology, the most specific child term should be used. We recommend using AmiGO 2 [22] (URL: http://amigo.geneontology.org/amigo) or QuickGO [23] (URL: https://www.ebi.ac.uk/QuickGO/) in order to easily explore the available terms while curating the DisProt evidence.

## Defining disorder-derived functions

It is possible to annotate functions directly derived from the disordered state of the protein by using the IDP ontology. Detailed information about each term stored in IDP ontology is available in the Ontology page of DisProt (URL: https://disprot.org/ontology).

### *Entropic chain*

Function directly arising from the lack of a stable structure. These entropic chain functions stem from the ability of the IDP to fluctuate between a large number of different conformational states. If known, the curator can be more specific about the type of entropic chain function by attaching terms under 'entropic chain' from the following: *flexible C-terminal tail, flexible N-terminal tail, flexible linker/spacer*.

### *Molecular recognition display site*

The flexibility of a post-translational modification (PTM) site is usually required to allow it to effectively fit into the active site of the modifying enzyme, therefore PTMs are usually associated with the presence of intrinsic disorder. Available terms under 'molecular recognition display site' include e.g. *glycosylation display site, limited proteolysis display site, phosphorylation display site*.

### *Self-regulatory activity*

Protein interaction in *cis* that auto-regulates the protein function or its assembly, e.g. *self-activation* and *self-inhibition*.

# The MIADE guidelines: Minimum Information About a Disorder Experiment

The IDP community has developed the Minimum Information About Disorder Experiments (MIADE) guidelines to unambiguously define an experimental setup used to study the structural aspects of IDPs or IDRs [24]. As extensively described in the article, the MIADE guidelines provide recommendations for data producers on how to describe the results of their IDP-related experiments, for biocurators on how to annotate the experimental data in manually curated resources, and for database developers on how to disseminate the data. In particular, MIADE increases the accuracy and accessibility of IDR annotations by providing information about experimental protocols, sample components, or sequence properties that might affect the interpretation of the experimental results.
Using MIADE, it is possible to objectively examine and compare experimental evidence from other sources that follow the same standard. Curators can add experimental-related information in DisProt by clicking the “Additional fields” buttons, which include **Sequence construct**, **Experimental conditions**, and **Experimental components**. For each annotation pertaining to this aspect, excerpts from the scientific article and ‘curators' comments can be added to provide more clarity to the annotation.

## Sequence construct information

The MIADE implementation allows defining differences of the protein-sequence described by the authors and UniProt protein sequence. Differences can arise from five factors that have been identified and described in the guidelines (Table 2 from [24]) and can alter the original sequence. In DisProt, it is possible to select one or more factors of the construct alterations and provide experimental details (Figure 5).

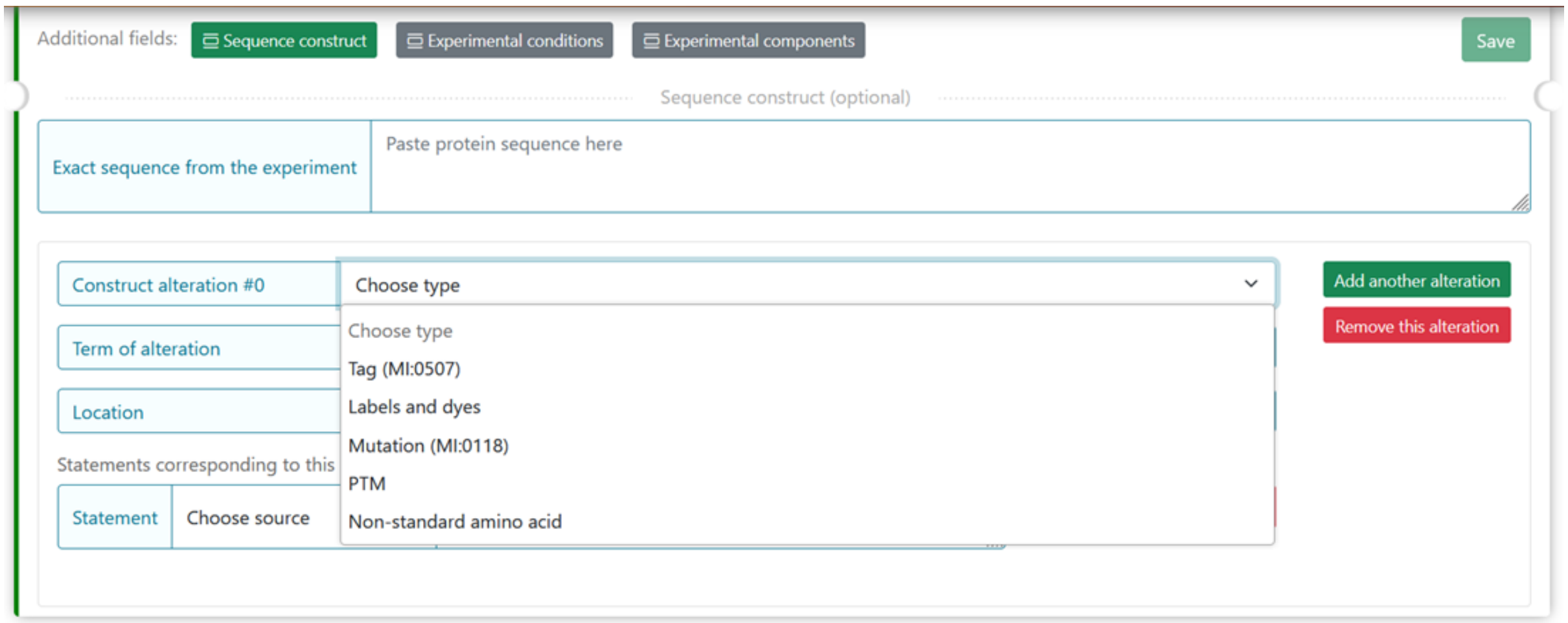

**Figure 5.** The figure shows the five selectable factors in DisProt to define any sequence differences reported in the experiment.

For each of the five factors, the construct alteration terms should be specified by choosing from the dropdown menu. Deviations from the canonical protein sequence should be described using appropriate ontologies:

- Tag and labels by the format standard of molecular interaction data, PSI-MI [25] (URL: https://www.ebi.ac.uk/ols/ontologies/mi).
- Mutations should be annotated using the HGVS nomenclature for the description of protein sequence variants [26] (URL: https://varnomen.hgvs.org/).
- Post-translational modifications (PTMs) and Non-standard amino acids should be indicated using the controlled vocabulary for the protein chemical modifications, PSI-MOD [27] (URL: https://www.ebi.ac.uk/ols/ontologies/mod).

For instance, the experimental construct can differ from the canonical sequence by the presence of a mutation. For the Cellular tumour antigen p53 (DisProt entry DP00086, URL: https://disprot.org/DP00086) (Figure 6), the sequence construct contains four non synonymous substitutions, which nature should be specified in DisProt by selecting from the dropdown menu.

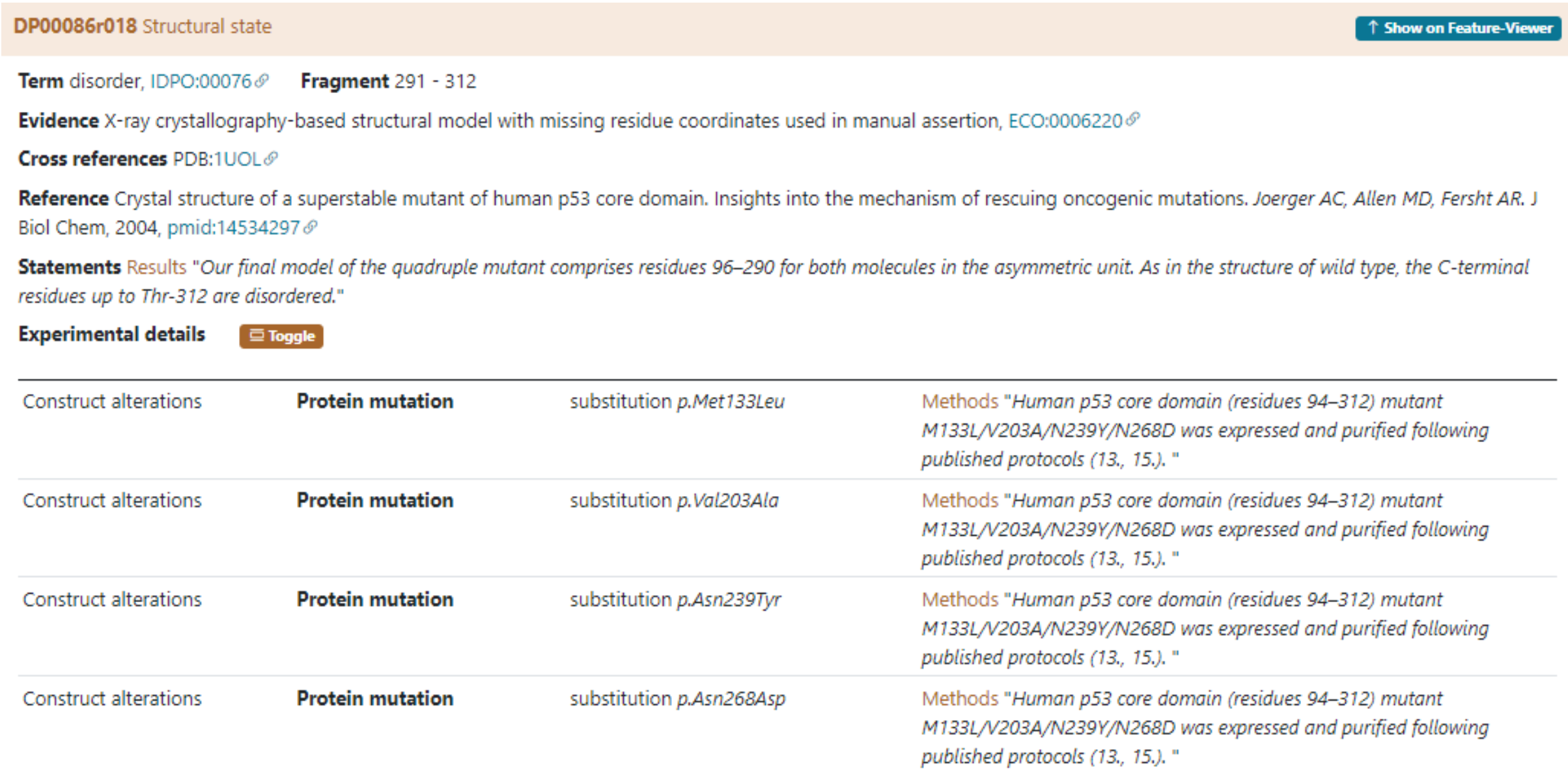

**Figure 6.** The figure shows the evidence of the disordered region 291-312 experimentally detected. The construct used by the authors contains four substitutions *M133L/V203A/N239Y/N268D* reported with the HGV nomenclature [26]. The curator has also added a statement extracted from the article in the Methods section to support the evidence related to the construct alteration.

## Definition of the experimental conditions

The experimental parameters of a given assessment can affect our comprehension of the biological significance of an experimental observation. Four parameter categories of the experimental setup for a sample, defined in the NCI Thesaurus OBO Edition controlled vocabulary (URL: https://ncit.nci.nih.gov/ncitbrowser/) (for details see Table 2 from [24]), should be specified in DisProt by choosing from the dropdown menu (Figure 7).
For each of the four properties absolute values (Units of Measurement Ontology) and deviations from the expected value (within normal range, increased, decreased, not specified or not relevant) can also be added.

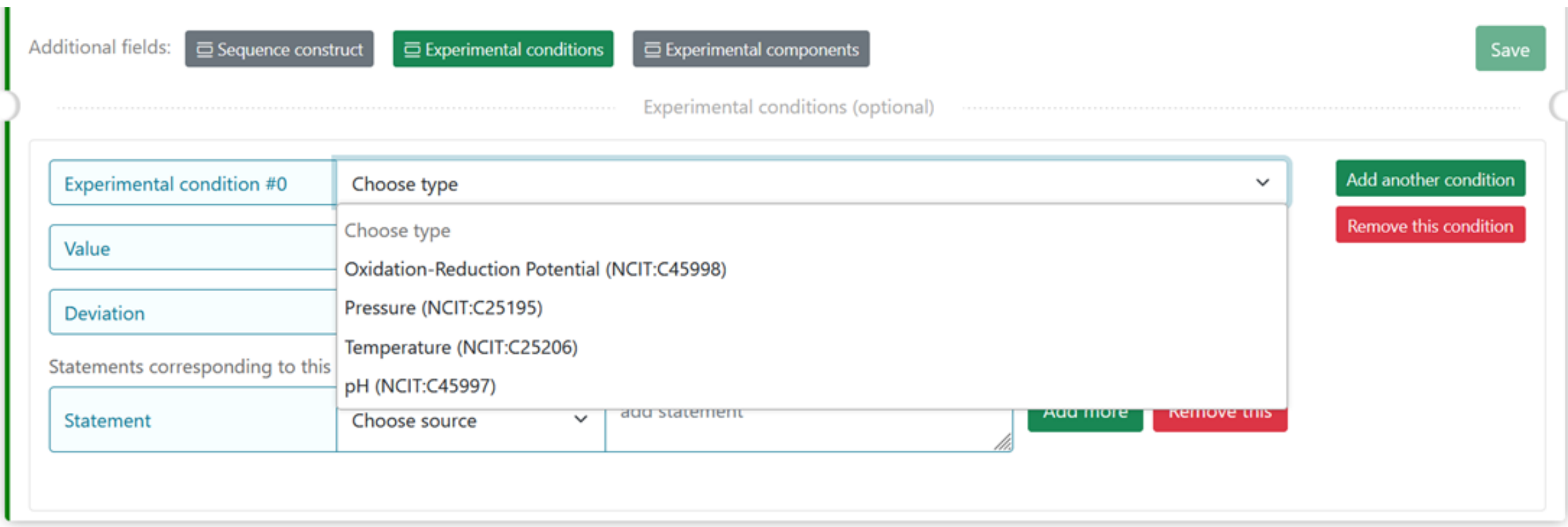

**Figure 7.** The figure shows the four selectable experimental parameters in DisProt to define any experimental setup for a sample in the study reported.

An example is the pH=5 reported in a crystallographic analysis of a soluble fragment of Hemagglutinin protein (DisProt entry DP03517, URL: https://disprot.org/DP03517) as reported in the details in Figure 8.

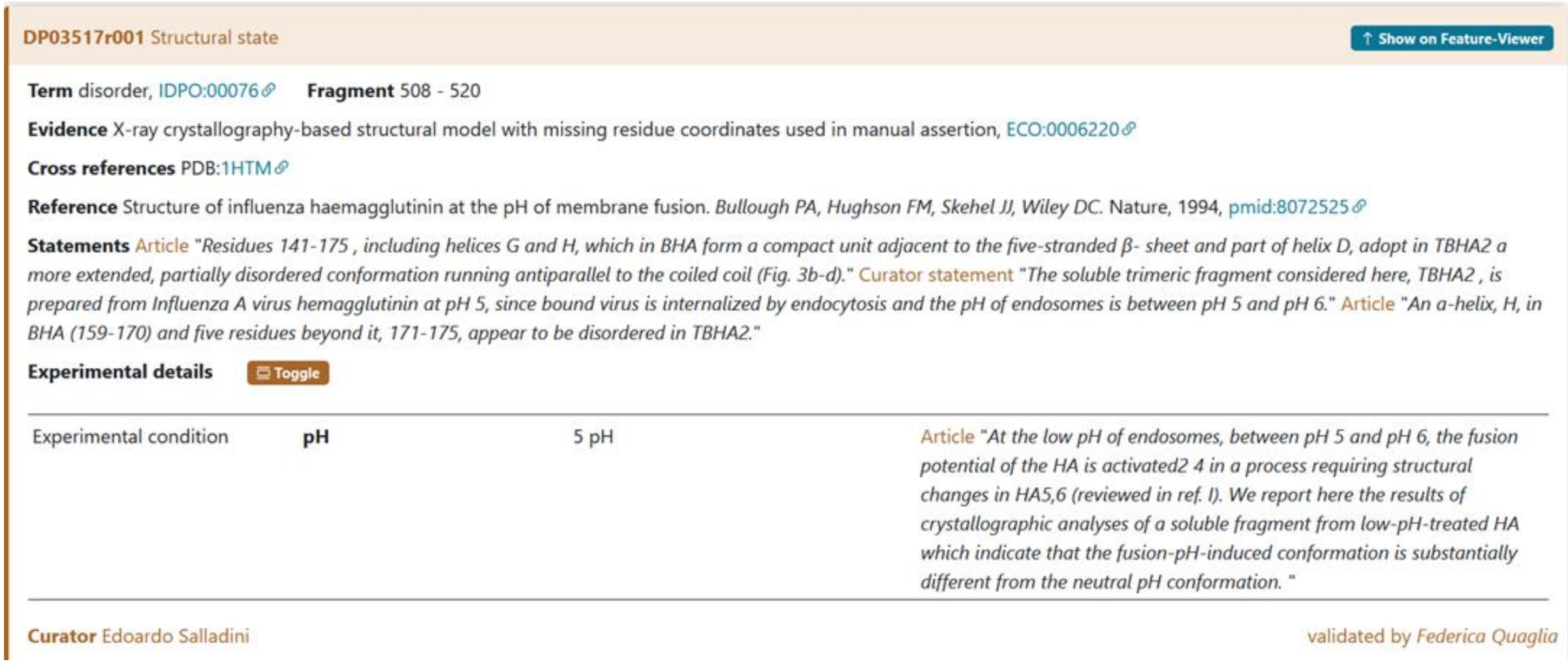

**Figure 8.** The figure shows the evidence of the disordered region 508-520 experimentally detected.The soluble fragment is prepared at pH=5 as mentioned in Experimental details. The curator has also added a statement extracted from the article to support the evidence related to the experimental conditions.

## Experimental components definition

Seven experimental sample components used by the authors during the characterization of an IDR, should be specified in DisProt by choosing in the dropdown menù as shown in Figure 8.
The interacting partners that may affect the correct interpretation of the experiment, have been defined in DisProt by the IDPO controlled vocabulary.
For each of the experimental components, the curator should specify the database to cross reference, based on the nature of the interaction partner (Table 2). However, the concentration and deviation can be added.

| **Sample component** | **Database** |
|---|---|
| In-cell experiment | Cellosaurus [28] |
| Interacting antibody | ABCD (Antibodies Chemically Defined) [29] |
| Interacting lipid | CAS Registry Number [30] |
| | ChEBI [31] |
| | PubChem [32] |
| Interacting membrane | CAS Registry Number [30] |
| | ChEBI [31] |
| Interacting Nucleic Acid | ENA |
| | RNAcentral [33] |
| Interacting Protein | CAS Registry Number [30] |
| | UniProt [19] |
| Interacting small molecule | CAS Registry Number [30] |
| | ChEBI [31] |
| | ChEMBL [34] |
| | PubChem [32] |

**Table 2.** The table shows parameters and databases from which to retrieve information regarding the experimental components that the curator should annotate in the specific section of a DisProt entry.

An example is the interacting proteins used by the authors during the characterization of the IDR in Cellular tumor antigen p53 (DisProt entry DP00086, URL: https://disprot.org/DP00086). The small interacting molecule with ChEBI ID:26710 is the sodium chloride (NaCl) as specified also in the statement by curators (Figure 9).

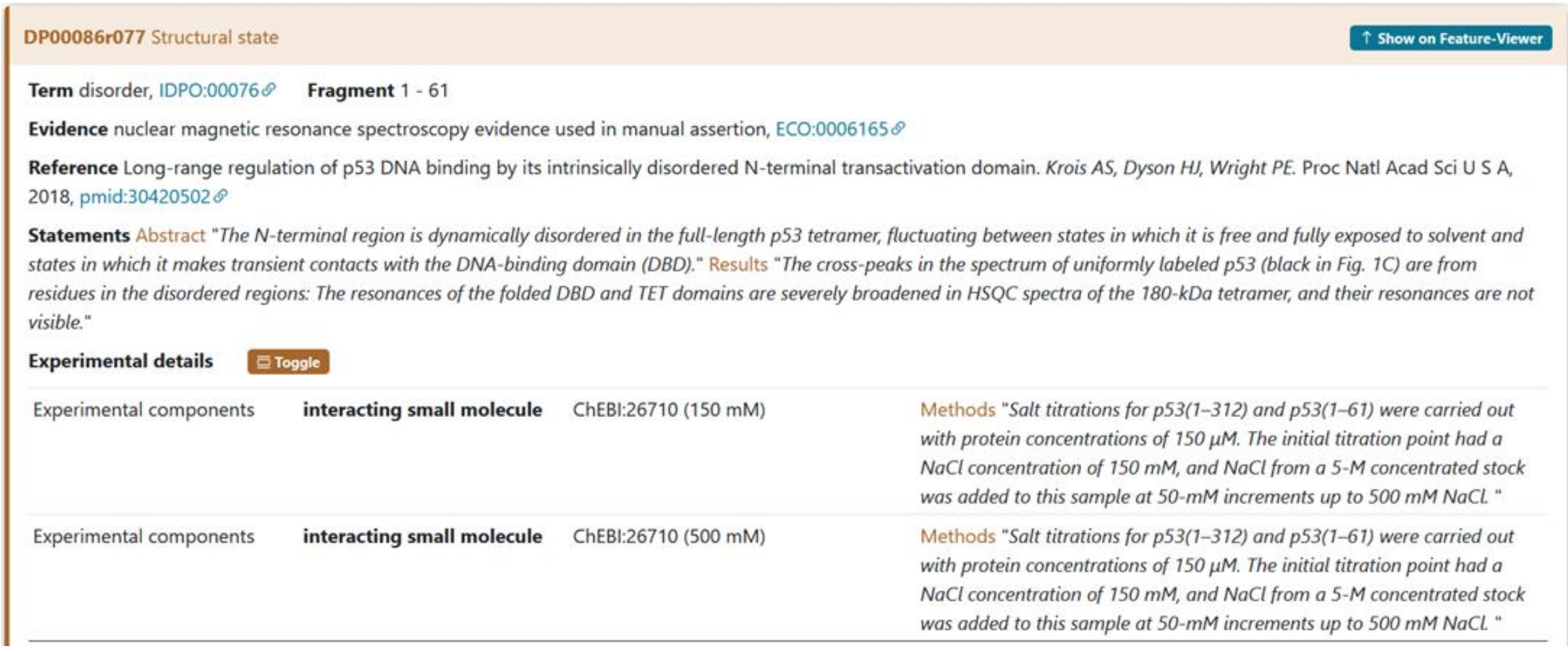

**Figure 9.** The figure shows the evidence of the disordered region 291-312 experimentally detected. The sample studied by the authors contains the NaCl molecule. The curator has also added a statement extracted from the article in the Methods section to support the evidence related to the interacting partners.

For further details and information about disorder-related experiments codified through MIADE in DisProt, we recommend curators to refer to the MIADE guidelines from Meszaros et al. [24].

## Thematic datasets

Since December 2020, DisProt has offered thematic datasets that are relevant to specific biological processes or organisms [2]. The construction of these datasets relies on collaborations established among experts in the respective fields. Consequently, curators have the option to focus their curation efforts on proteins that are linked to a thematic dataset and contribute to their enrichment. They also have the opportunity to propose the creation of new thematic datasets.

The protein selection provided to curators for dataset construction or enrichment comes from reliable data sources, including curated databases specialising in particular topics. For example, "Cancer-related proteins' dataset, has been constructed with proteins included in COSMIC [35]. Similarly, for the 'NDDs-related proteins' dataset, proteins associated with Neurodevelopmental Disorders (NDDs) were selected from resources such as SFARI [36] and SysNDD [37].
Regarding the practical aspect, during the annotation process, the curator should also add a specific tag to the protein if it belongs to an available dataset (Figure 10).

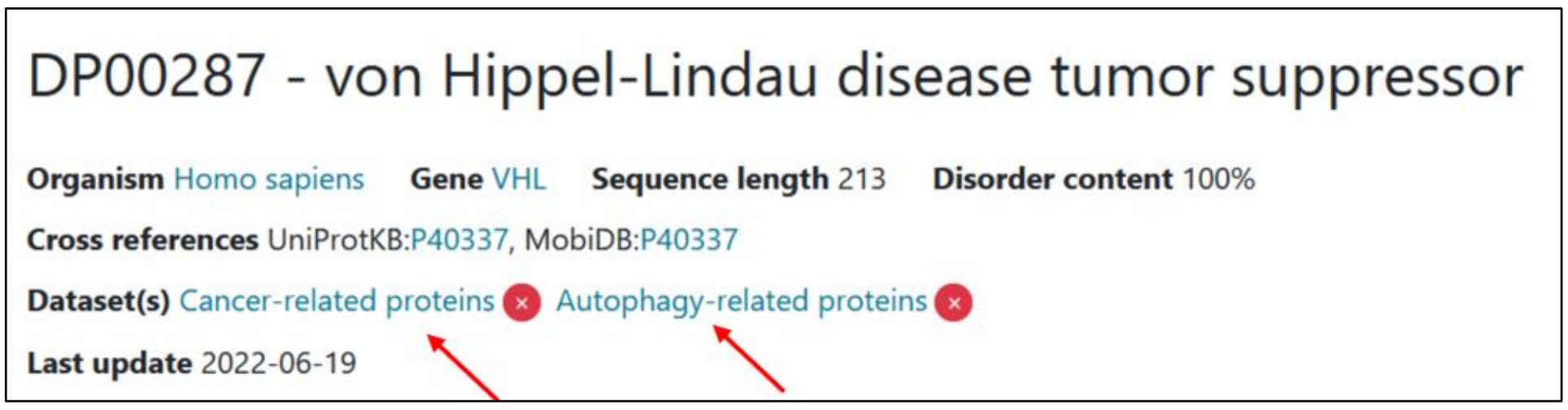

**Figure 10.** Example of a protein entry associated with two thematic datasets, with the two added tags pointed out by red arrows.

# IDP literature curation use cases

The following sections provide guidance on using data available within scientific articles to create an entry or new evidence for curating disordered proteins and regions in DisProt. Figure 11 shows an overview of the specific steps for adding information about an IDP/IDR for the ATG8-interacting protein 2 (ATI2) in DisProt.

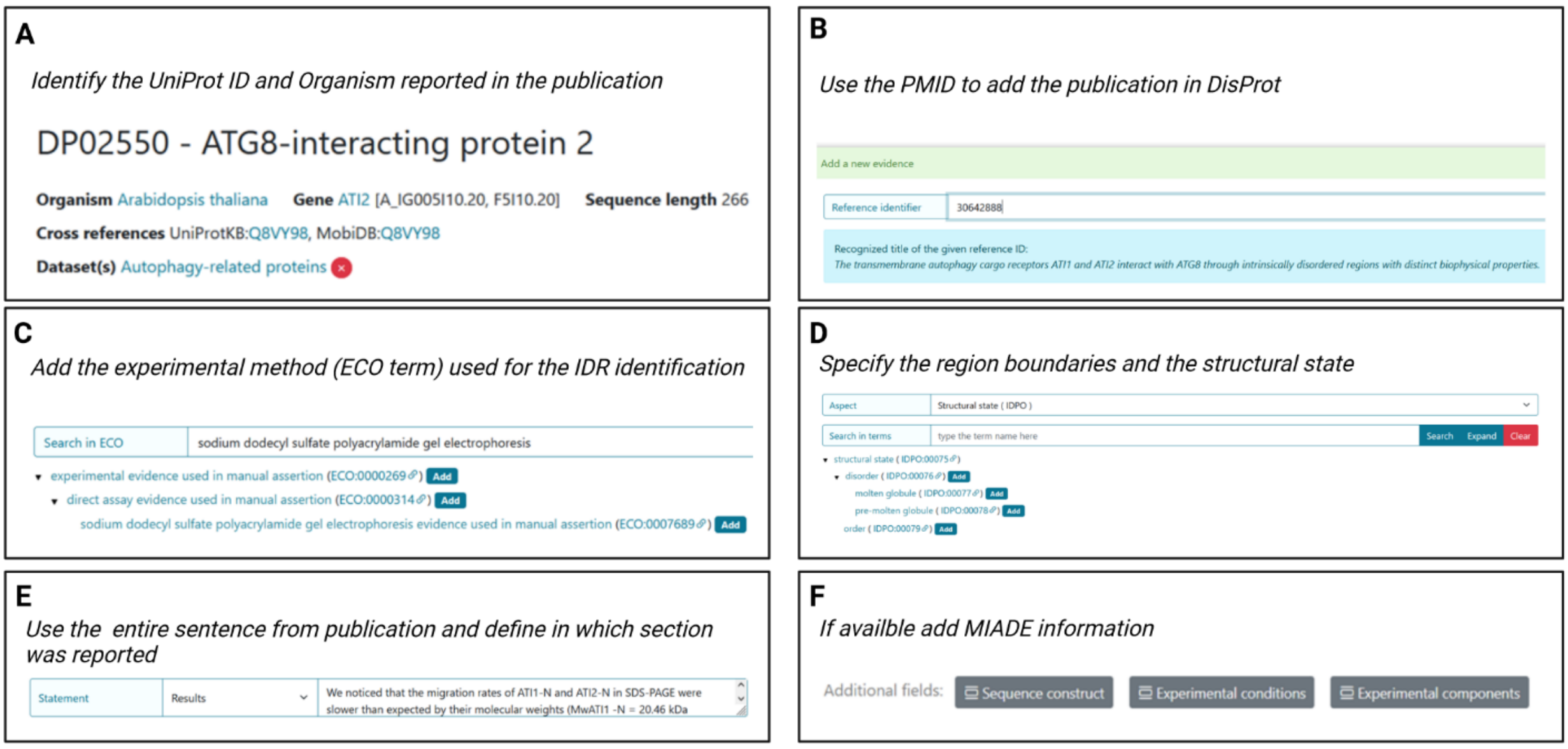

**Figure 11. DisProt representation of the steps performed during the manual curation of ATG8-interacting protein. A**. A UniProt identifier should be used for the entry creation. **B.** Include the PMID to cite the publication as the source of information. The title is automatically retrieved. **C.** The method employed for the assessment should be chosen from the list of available ECO terms. **D.** In accordance with the UniProt sequence, the curator should report the "start" and "end" positions of the IDR. From the drop-down menu choose the right aspect: structural state (IDPO), structural transition (IDPO) or Function (IDPO or GO) and select the specific term. **E.** As support of the evidence and annotation, curators are required to add statements from the publication. Curators should copy and paste the original sentences as they appear in the article, while also specifying the exact section where it can be located (eg. Results). **F.** Additional information corresponding to sequence construct, experimental conditions and /or experimental components can be added if suitable.

## ATG8-interacting protein 2 (ATI2)

To elucidate the disorder status and functional role of the N-terminus in the ATI2 protein, the authors of the article entitled "*The transmembrane autophagy cargo receptors ATI1 and ATI2 interact with*

*ATG8 through intrinsically disordered regions with distinct biophysical properties*” published in The Biochemical Journal [38], conducted various experiments and engaged in a comprehensive discussion regarding the intrinsically disordered regions (IDR) and their significance within the ATI2 protein. The corresponding protein entry is already annotated in DisProt (DisProt entry DP02550, URL: https://disprot.org/DP02550).

## Finding the UniProt ID for ATI2 protein

The first step in curating the referenced publication is to identify the UniProt ID for one of the two proteins mentioned in the paper, ATI2, paying attention to the organism considered: *Arabidopsis thaliana*. The authors refer to the protein by its official symbol name, ATI2, and the alternative name, At4G00355, corresponding to the Q8VY98 as UniProt ACC.

## Identifying intrinsic disorder structural state of ATI2

The disorder status assessment of the ATI2 N-terminus is performed by five experiments (Table 3). The IDPO ontology available in DisProt should be used to indicate the nature of the N-terminal region of ATI2 protein with the specific term IDPO:00076**:** “*disorder*”, according to what is stated in the publication. In this particular case, the authors are very clear in defining the region of the ATI2 protein as disordered, both in the title and in the text. Hence, the curator ought to choose excerpts from the publication that unequivocally endorse and elucidate the disorder of the region, incorporating them as unmodified statements through the process of copying and pasting.

In this case, there will be five distinct pieces of evidence for the disorder status of the region ranging 1-193 residues, each of them supported by different methods and supported by the corresponding statement from the publication (See Table 3).

| **Data to be annotated for the Structural state** | **Description** |
|---|---|
| Scientific publication | *The transmembrane autophagy cargo receptors ATI1 and ATI2 interact with ATG8 through intrinsically disordered regions with distinct biophysical properties* [38] |
| Organism | *Arabidopsis thaliana* |
| Protein | ATG8-interacting protein 2 (ATI2) |
| UniProt ID | Q8VY98 |
| Region boundaries | 1 - 193 |
| Experimental methods for the IDR assessment | SDS-PAGE, size-exclusion chromatography, far-UV circular dichroism, NMR spectroscopy, temperature-induced protein unfolding |
| Aspect of the annotation and term | Disorder |
| Publication statement | "*We noticed that the migration rates of ATI1-N and ATI2-N in SDS-PAGE were slower than expected by their molecular weights (MwATI1 -N = 20.46 kDa; MwATI2-N = 21.02 kDa) (Fig. 3A). This simple observation represents a first indication that ATI1 -N and ATI2-N may be intrinsically disordered, because IDRs are typically depleted in hydrophobic residues, and, consequently, tend to bind less SDS, explaining their abnormally slow mobility in* |
| | *SDS-PAGE [37].* " (SDS-PAGE) |

**Table 3. The table contains data extracted from the publication [38] used for characterising the ATI2 protein's disordered state.** The region flexibility (1-193) was verified by five different techniques, all of which has been included in DisProt as five separate pieces of evidence, associated with each experimental method and supported by a corresponding statement from the publication. One statement as an example was reported in this table.

### Defining the IDR functions of the ATI2 AIM-motif through Gene Ontology terms

By using two different methods (Table 4), the authors have further identified and characterised the presence of an ATG8-interacting motif (AIM) located at position 14-17 (residues WEVV) of ATI2, within the intrinsically disordered region (1-193). This feature should be annotated as a function employing the Gene Ontology term "*Protein binding*" (GO:0005515) where the binding partner, the Autophagy-related protein 8f (ATG8F), should be also specify by its UniProt ID (Q8VYK7).
Additionally, this region is implicated in selective autophagy, as demonstrated by two techniques (Table 4). This functional attribute can also be annotated using the Gene Ontology term for "*selective autophagy*" (GO:0061912).

- *Molecular function - protein binding, GO:0005515*
- *Biological process - selective autophagy, GO:0061912*

Note that in the DisProt annotations related to these features, the boundaries (13-18) differ from those reported in the publication and in the UniProt canonical isoform. This divergence results from functional annotation requirements for which a **minimum of 5 residues** are necessary for the annotation in DisProt. Consequently, the curator should extend the boundaries by one amino acid.

| **Data to be annotated for the Disorder Function** | **Description** |
|---|---|
| Scientific publication | *The transmembrane autophagy cargo receptors ATI1 and ATI2 interact with ATG8 through intrinsically disordered regions with distinct biophysical properties [38]* |
| Protein | ATG8-interacting protein 2 (ATI2) |
| Organism | *Arabidopsis thaliana* |
| UniProt ID | Q8VY98 |
| Experimental methods for IDR functional assessment | Heat treatment, yeast two-hybrid assays, NMR spectroscopy |
| Region boundaries | 13-18 |
| Curator statement | "*Region including the N-terminal AIM motif (WEVV) at position 14-* |
| | *17.*" |
| MIADE | Experimental details:<br>● Construct alterations<br>● Substitution: *p.Trp14Ala, p.Val17Ala* |

**Table 4**. The table contains all data extracted from the publication [38] useful for characterising the disorder function of ATI2 protein. Two different functions: protein binding and selective autophagy were demonstrated by the authors and supported in DisProt by corresponding statement and cross reference. One statement as an example was reported in this table.

## RAF proto-oncogene serine/threonine-protein kinase (RAF1)

The following example describes a RAF1 protein for which information about disordered state is derived from two publications "*Synergistic binding of the phosphorylated S233- and S259-binding sites of C-RAF to one 14-3-3ζ dimer* [39]" published in Journal of Molecular Biology and *"Stabilization of physical RAF/14-3-3 interaction by cotylenin A as treatment strategy for RAS mutant cancers* [40]*"* published in Chem Biol. The corresponding protein entry is already annotated in DisProt (DisProt entry DP00171, URL: https://disprot.org/DP00171).
The specific sections will be informative for curating the disordered state of a phosphorylated peptide when it is in complex with a protein partner and how to support the disordered state (eg. Statement) and experimental conditions (MIADE). It should be noted that in both publications the structure of the di-phosphorylated RAF1 peptide in complex with the 14-3-3ζ shows an intrinsically disordered about 20 amino acids region.

### Finding a UniProt identifier for RAF1 protein

The UniProt identifier was obtained by searching UniProt using the protein name, RAF1, and the organism of origin, *Homo sapiens*, as stated by the authors in the publication. The amino acid sequence of the UniProt entry P04049 (URL:https://www.uniprot.org/uniprotkb/P04049/entry) was compared to the amino acid sequence of the synthetic RAF1 construct used in the publication: *QHRYSTPHAFTFNTSSPSSEGSLSQRQRSTSTPNVH* (shown in the Figure 1) to ensure its identity.

### Defining the experiments used to characterise RAF1

To determine the structural basis of the interaction between RAF1 and the 14-3-3ζ protein the authors performed X-ray diffraction analysis of the crystallised complex. Even though the ECO term "*X-ray crystallography evidence used in manual assertion"* (*ECO:0005670*) is acceptable to describe the technique, the more specific child term "*X-ray crystallography-based structural model with missing residue coordinates used in manual assertion"* (*ECO:0006220)* is more suitable and should be selected to describe the experimental procedure in which the authors based their observations about RAF1 structural state. After adding both PMIDs, the system will automatically provide the PDBs code as a cross-references. In this case, since the inserted technique is X-ray and the amino acid residues mentioned by the authors are missing, PDB:4FJ3 and PDB:4IHL can be added by the curators.

### Identifying intrinsic disorder structural state of RAF1

Molzan M. et al. [39] co-crystallized the 14-3-3ζ protein with a synthetic di-phosphorylated peptide of 36 amino acids that corresponds to the 229-264 residues of RAF1. While it was possible to crystallise 14 residues of RAF1, associated with the 14-3-3ζ interacting region, it was not possible to trace a 20-amino- acid stretch between the phosphorylated sites (Ser 233 and Ser 259).
The lack of electron density in the region 236-255 indicates significant flexibility, signifying an intrinsically disordered region, as defined by the IDPontology term "*dis*order" (IDPO:00076). In the work of 2013 by Molzan M. et al. [39], the 14-3-3ζ protein was crystallised in the presence of the same synthetic peptide (RAF1) and the natural product Cotylenin A. In this case the disordered stretch was shorter than the previous one (238-254) and the authors did not explicitly restate its disordered nature. Nevertheless, based on the observed missing electron density in the PDB file, the curator can independently confirm the disorder status of this region (Figure 6).

| **Data to be annotated for the Structural state** | **Description** | **Description** |
|---|---|---|
| Scientific publication | *Synergistic binding of the phosphorylated S233- and S259-binding sites of C-RAF to one 14-3-3ζ dimer.* | *Stabilization of physical RAF/14-3-3 interaction by cotylenin A as treatment strategy for RAS mutant cancers.* |
| Organism | *Homo Sapiens* | *Homo Sapiens* |
| Protein | RAF proto-oncogene serine/threonine-protein kinase | RAF proto-oncogene serine/threonine-protein kinase |
| UniProt ID | P04049 | P04049 |
| Experimental methods for IDR assessment | X-ray crystallography | X-ray crystallography |
| Region boundaries | 236-255 | 238-254 |
| Statement | From Article | By curator |
| MIADE | • Construct alterations<br>• Experimental components | • Construct alterations<br>• Experimental components |

**Table 5.** The region flexibility (238-254) was verified by X-ray technique, included in DisProt and supported by a corresponding statement from the publication. One statement as an example was reported in this table.

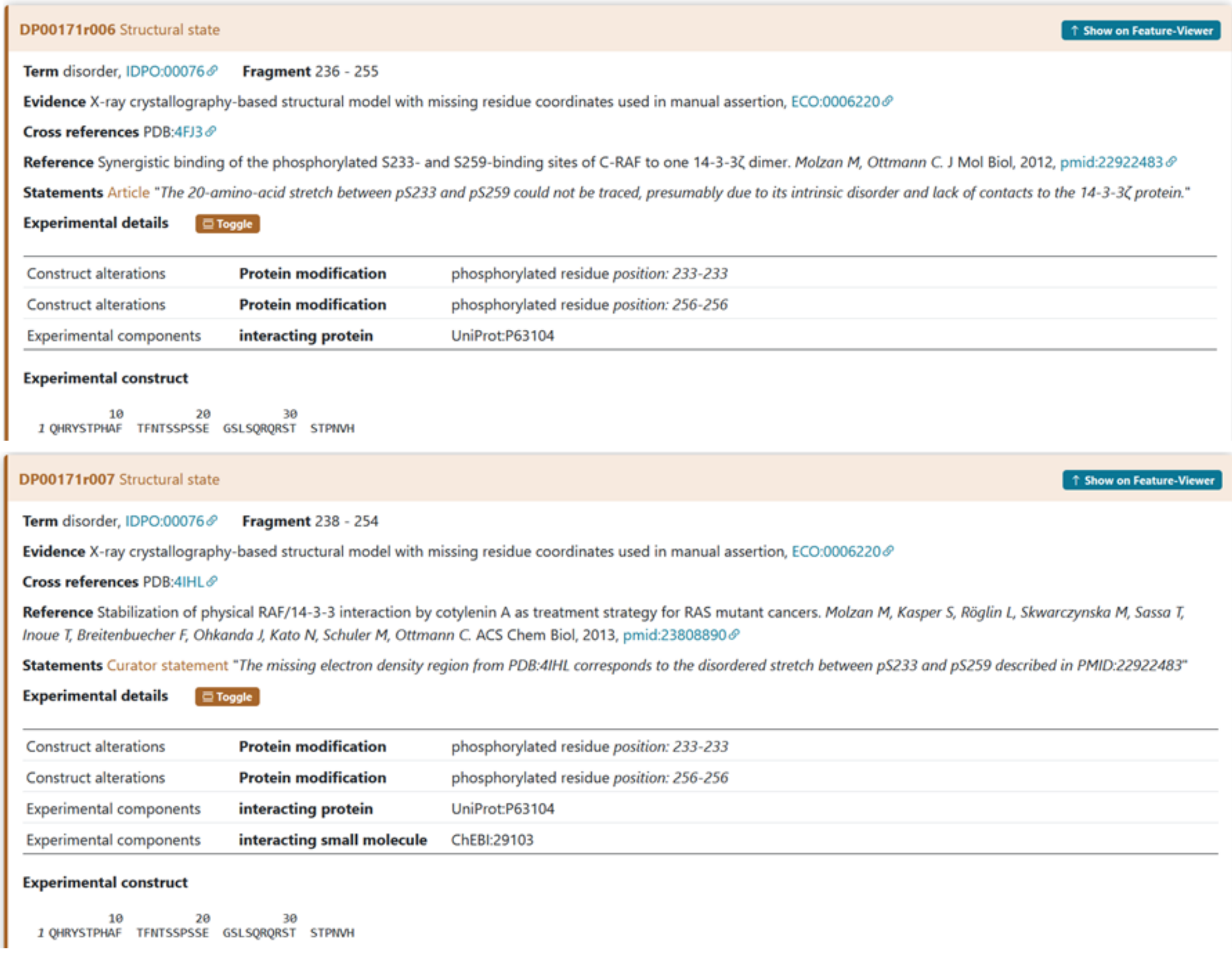

**Figure 13.** Two pieces of evidence of the structural state for the RAF1 intrinsically disordered region.

### Defining MIADE specifications

As stated previously, modifications of the amino acid sequence or the presence of molecular partners, could affect the result of a given experiment and should be taken into consideration for its correct interpretation.

The annotation regarding the X-ray evidence of disorder region in RAF1 should also report the presence of the interacting protein 14-3-3ζ (UniProt ACC P63104), and the presence of the PTM modifications of the serine residues 233 and 256 (Figure 13). This information can be found in the PDB deposited structure and in the method section of the publication.

# Training courses: future prospective

The training activity in DisProt is the foundation for the curation, implementation, and expansion of the database. Every curator, before receiving their account for curation in DisProt, should complete a course available on the ELIXIR eLearning platform (URL: https://elixir.mf.uni-lj.si/enrol/index.php?id=91). The course, currently available in English and Spanish, provides curators with most relevant information to start their biocuration activity.

Other curation materials are available as webinars and curation manuals. In particular, in ELIXIR training Portal - TESS [41], it is possible to find a beginner and intermediate level course describing the DisProt resource (URL: https://tess.elixir-europe.org/search?q=disprot). A team of experts will be engaged to video-record training sessions focused on experimental methods, making them accessible to anyone wishing to deepen their knowledge and interpretation of disorder-related data experimentally studied in the literature.

## Recognition and Accreditation in DisProt

One of the most important policies of DisProt is to consistently reward the effort and meticulous work of expert and volunteer curators. In this context, DisProt was one of the first databases to be integrated into APICURON, a platform developed with the purpose of accrediting the work of biocurators based on the concept of gamification (URL: https://apicuron.org/) [42]. The recognition of a DisProt biocurator's activity takes into account the effort, accuracy, and data quality added to the DisProt database. The terms and scores that accredit the curator's activity are based on the importance and the purpose of encouraging further exploration of that data in DisProt.

## Submitting an annotation of intrinsic disorder to DisProt

Sharing novel insights on intrinsically disordered proteins (IDPs) and regions (IDRs) into the DisProt database was streamlined through a dedicated submission form, ensuring the integration of manually curated literature findings. The submission form is accessible through the DisProt website (URL: https://disprot.org/biocuration), in which users find all necessary fields for accurate annotation. Initiating the process, users provide the UniProt accession number (*UniProt ACC*) for precise cross-referencing. If available, the DisProt identifier can also be entered in order to enhance the linkage. Essential contact information, including *Email*, *Full name*, and *ORCID*, guarantees proper attribution to the author of the submission. Contributors are required to add, for each new submitted evidence, the PubMed reference identifier of the peer-reviewed scientific publication describing the IDP/IDR, as well as the characterised region , the experimental technique and the disorder aspect described. Contributors also have the chance to fill out additional fields, such as *Cross reference* and *Statement*, to contextualise the annotation. The submission process accommodates for additional information in the *Comment* section, fostering clarity.

## Conclusions

Intrinsically disordered proteins (IDPs) and regions (IDRs) represent essential components of the proteome, playing diverse and pivotal roles across biological processes and functions. Biocuration ensures an accurate and standardised representation of the inherent complexity of IDPs and IDRs properties, functions, and interactions, thus facilitating a comprehensive understanding of their different contributions to cellular dynamics.

The biocuration of these proteins from peer-reviewed literature forms the basis of knowledge enrichment within the DisProt database.

The process of curation is well defined and follows a structured approach to help curators carefully extract relevant information from scientific publications. This includes precise delineation of protein boundaries, thorough documentation of experimental methods, and meticulous recording of disorder-related details.

The DisProt database integrates ontologies like the Intrinsically Disordered Proteins Ontology (IDPO), Gene Ontology (GO), and the Evidence and Conclusion Ontology (ECO) to provide structured data annotations. These ontologies bring consistency to the classification of IDP/IDR attributes, ranging from structural states and transitions to molecular functions and biological processes, resulting in coherent cross-referencing and interpretation.

Moreover, incorporating the principles of the Minimum Information About a Disorder Experiment (MIADE) guidelines, we elevate our curation practices. MIADE guidelines serve as the foundation upon which we build our data representation. They ensure that every piece of relevant information is captured and presented with the utmost precision, resulting in comprehensive and trustworthy data sets.

This guideline, designed for both biocurators and external users, provides a step-by-step guide for the systematic and thorough curation of IDPs and IDRs within the DisProt framework.

By addressing every aspect of the curation process and by providing practical examples of well-known IDPs, this guideline allows curators and users to explore the rigorous curation best practices that ensure the maintenance of DisProt's high standards of data accuracy and reliability.

## Acknowledgements

This publication is based on work from a project that has received funding from the European Union’s Horizon 2020 research and innovation programme under grant agreement No 778247 (H2020 Marie Sklodowska Curie Actions RISE “IDPfun”). This work was also funded by ELIXIR, the research infrastructure for life-science data, and the ELIXIR IDP Community and by the European Union through NextGenerationEU, PNRR project - "ELIXIR x NextGenerationIT: Consolidamento dell'Infrastruttura Italiana per i Dati Omici e la Bioinformatica - ElixirxNextGenIT" - IR0000010; and for MCA the PNRR - CN - G.T.RNA SP. 7 project - National Center for Gene Therapy and Drugs based on RNA Technology (CN3) - CN00000041. Additionally, this work was partially funded by the Italian Ministry of University and Research (MIUR), through NextGenerationEU, PRIN project “PLANS: Proximity Ligation And Nanopore Sequencing for the characterization of native RNA-protein interactions” (2022W93FTW).

## References

1. Romero P, Obradovic Z, Kissinger C et al. Thousands of proteins likely to have long disordered regions. *Pac Symp Biocomput Pac Symp Biocomput* 1998.
2. Aspromonte MC, Nugnes M. Victoria et al. DisProt in 2024: improving function annotation of intrinsically disordered proteins. *Nucleic Acids Res* in press.
3. Quaglia F, Mészáros B, Salladini E *et al.* DisProt in 2022: improved quality and accessibility of protein intrinsic disorder annotation. *Nucleic Acids Res* 2021;**50**:D480–7.
4. Hatos A, Hajdu-Soltész B, Monzon AM *et al.* DisProt: intrinsic protein disorder annotation in 2020. *Nucleic Acids Res* 2020, DOI: 10.1093/nar/gkz975.
5. Piovesan D, Tabaro F, Mičetić I *et al.* DisProt 7.0: a major update of the database of disordered proteins. *Nucleic Acids Res* 2017;**45**:D219–27.
6. van der Lee R, Buljan M, Lang B *et al.* Classification of Intrinsically Disordered Regions and Proteins. *Chem Rev* 2014;**114**:6589–631.
7. Wright PE, Dyson HJ. Intrinsically Disordered Proteins in Cellular Signaling and Regulation. *Nat Rev Mol Cell Biol* 2015;**16**:18–29.
8. Fiorini N, Lipman DJ, Lu Z. Towards PubMed 2.0. *eLife* 2017;**6**:e28801.
9. Europe PMC: a full-text literature database for the life sciences and platform for innovation. *Nucleic Acids Res* 2015;**43**:D1042–8.
10. Aleksander SA, Balhoff J, Carbon S *et al.* The Gene Ontology knowledgebase in 2023. *Genetics* 2023;**224**:iyad031.
11. Velankar S, Best C, Beuth B *et al.* PDBe: Protein Data Bank in Europe. *Nucleic Acids Res* 2010;**38**:D308–17.
12. Hoch JC, Baskaran K, Burr H *et al.* Biological Magnetic Resonance Data Bank. *Nucleic Acids Res* 2023;**51**:D368–76.
13. Valentini E, Kikhney AG, Previtali G *et al.* SASBDB, a repository for biological small-angle scattering data. *Nucleic Acids Res* 2015;**43**:D357–63.
14. Lawson CL, Baker ML, Best C *et al.* EMDataBank.org: unified data resource for CryoEM. *Nucleic Acids Res* 2011;**39**:D456–64.
15. Whitmore L, Miles AJ, Mavridis L *et al.* PCDDB: new developments at the Protein Circular Dichroism Data Bank. *Nucleic Acids Res* 2017;**45**:D303–7.
16. Mészáros B, Erdős G, Szabó B *et al.* PhaSePro: the database of proteins driving liquid-liquid phase separation. *Nucleic Acids Res* 2020;**48**:D360–7.
17. Varadi M, De Baets G, Vranken WF *et al.* AmyPro: a database of proteins with validated amyloidogenic regions. *Nucleic Acids Res* 2018;**46**:D387–92.
18. Kumar M, Michael S, Alvarado-Valverde J *et al.* The Eukaryotic Linear Motif resource: 2022 release. *Nucleic Acids Res* 2022;**50**:D497–508.
19. The UniProt Consortium. UniProt: the Universal Protein Knowledgebase in 2023. *Nucleic*

*Acids Res* 2023;**51**:D523–31.
20. Piovesan D, Del Conte A, Clementel D *et al.* MobiDB: 10 years of intrinsically disordered proteins. *Nucleic Acids Res* 2023;**51**:D438–44.
21. Kuloğlu ES, McCaslin DR, Kitabwalla M *et al.* Monomeric Solution Structure of the Prototypical ‘C’ Chemokine Lymphotactin,. *Biochemistry* 2001;**40**:12486–96.
22. Carbon S, Ireland A, Mungall CJ *et al.* AmiGO: online access to ontology and annotation data. *Bioinformatics* 2009;**25**:288–9.
23. Binns D, Dimmer E, Huntley R *et al.* QuickGO: a web-based tool for Gene Ontology searching. *Bioinformatics* 2009;**25**:3045–6.
24. Mészáros B, Hatos A, Palopoli N *et al.* MIADE metadata guidelines: Minimum Information About a Disorder Experiment. 2022:2022.07.12.495092.
25. Sivade (Dumousseau) M, Alonso-López D, Ammari M *et al.* Encompassing new use cases - level 3.0 of the HUPO-PSI format for molecular interactions. *BMC Bioinformatics* 2018;**19**:134.
26. den Dunnen JT, Dalgleish R, Maglott DR *et al.* HGVS Recommendations for the Description of Sequence Variants: 2016 Update. *Hum Mutat* 2016;**37**:564–9.
27. Montecchi-Palazzi L, Beavis R, Binz P-A *et al.* The PSI-MOD community standard for representation of protein modification data. *Nat Biotechnol* 2008;**26**:864–6.
28. Bairoch A. The Cellosaurus, a Cell-Line Knowledge Resource. *J Biomol Tech JBT* 2018;**29**:25–38.
29. Lima WC, Gasteiger E, Marcatili P *et al.* The ABCD database: a repository for chemically defined antibodies. *Nucleic Acids Res* 2020;**48**:D261–4.
30. CAS REGISTRY | CAS.
31. Hastings J, de Matos P, Dekker A *et al.* The ChEBI reference database and ontology for biologically relevant chemistry: enhancements for 2013. *Nucleic Acids Res* 2013;**41**:D456–63.
32. Kim S, Chen J, Cheng T *et al.* PubChem 2023 update. *Nucleic Acids Res* 2022;**51**:D1373–80.
33. RNAcentral 2021: secondary structure integration, improved sequence search and new member databases. *Nucleic Acids Res* 2020;**49**:D212–20.
34. Gaulton A, Hersey A, Nowotka M *et al.* The ChEMBL database in 2017. *Nucleic Acids Res* 2016, DOI: 10.1093/nar/gkw1074.
35. Cosmic. COSMIC - Catalogue of Somatic Mutations in Cancer.
36. SFARI Gene - Welcome. *SFARI Gene*.
37. Home | SysNDD - The expert curated database of gene disease relationships in neurodevelopmental disorders.
38. The transmembrane autophagy cargo receptors ATI1 and ATI2 interact with ATG8 through intrinsically disordered regions with distinct biophysical properties | Biochemical Journal | Portland Press.
39. Molzan M, Ottmann C. Synergistic Binding of the Phosphorylated S233- and S259-Binding Sites of C-RAF to One 14-3-3ζ Dimer. *J Mol Biol* 2012;**423**:486–95.
40. Stabilization of Physical RAF/14-3-3 Interaction by Cotylenin A as Treatment Strategy for RAS Mutant Cancers | ACS Chemical Biology.
41. Beard N, Bacall F, Nenadic A *et al.* TeSS: a platform for discovering life-science training opportunities. *Bioinformatics* 2020;**36**:3290–1.
42. Hatos A, Quaglia F, Piovesan D *et al.* APICURON: a database to credit and acknowledge the work of biocurators. *Database J Biol Databases Curation* 2021;**2021**:baab019.